\documentclass[12pt]{article}
\usepackage{epsfig}

\newcommand{\be}{\begin{eqnarray}}
\newcommand{\ee}{\end{eqnarray}}

\begin{document}

\bibliographystyle{unsrt}
\footskip 1.0cm

\thispagestyle{empty}

\begin{flushright}
INT--PUB--05--32
\end{flushright}
\vspace{0.6in}

\begin{center}
\large {\bf {Ultra-High Energy Neutrino-Nucleon Scattering \\
and \\
Parton Distributions at Small $x$}}\\

\vspace{1.1in}

Ernest M. Henley$^{a,b}$ and Jamal Jalilian-Marian$^{b}$\\
\vspace{0.2in}

{\small \em $^a$Department of Physics, University of Washington, 
Seattle, WA 98195-1560\\
 $^b$ Institute for Nuclear Theory, University of Washington, 
Seattle, WA 98195-1550
}
\medskip

\bigskip
\begin{abstract} 

\noindent
The cross section for ultra-high energy neutrino-nucleon scattering   
is very sensitive to the parton distributions at very small values of 
Bjorken x ($x \leq 10^{-4})$. We numerically investigate the effects 
of modifying the behavior of the gluon distribution function at very 
small $x$ in the DGLAP evolution equation. We then use the Color Glass 
Condensate formalism to calculate the neutrino-nucleon cross section 
at ultra-high energies and compare the result with those based on 
modification of DGLAP evolution equation.

\end{abstract}
\end{center}
\medskip

\newpage

\section {Introduction}

Neutrinos of ultra-high energies ($E_\nu \geq 10^{7} GeV$) have been a puzzle for some time. 
One of the prime questions is where they come from, especially those above the GZK limit 
\cite {GZK}. Possible sources include decays of super massive particles (dark matter?), 
acceleration in active galactic nuclei, and supernovae explosions \cite {2}.  Another 
question of interest is the cross section for the scattering of these ultra-high energy 
neutrinos with nucleons. Here, part of the interest stems from the fact that if the cross 
section increases sufficiently rapidly, then the unitarity limit may be reached \cite {reno}. 
Another interest is what one can learn about the very small x parton distributions, since 
the energy dependence of the inclusive cross section is very sensitive to them. 

The cross sections for scattering of neutrinos on  nucleons at ultra-high energies are 
dominated by the gluons in the nucleon while the contribution of sea quarks 
is suppressed by $\alpha_s$ since they come from gluon splitting via  
$g \rightarrow q \bar{q}$. For $ x \leq 10^{-2}$ the gluon distribution function of a 
nucleon is known to grow fast 
\cite{HERA} with increasing $Q^2$ (virtuality of the gauge boson exchanged) and 
decreasing $x$ as $(1/x)^\beta $, with beta less than 1. This implies 
that the structure functions, e.g., $F_2$, in deep inelastic scattering will also 
increase, which would in turn mean a fast increase of the neutrino-nucleon total cross 
section. This fast growth of the total cross section can not continue indefinitely since
it would violate unitarity (the Froissart bound). The parton (gluon) phase space density 
(number of partons per unit area and rapidity) is expected to be very high at very small 
Bjorken $x$ which would lead to an overlap in transverse space and recombination of gluons 
which in turn could lead to saturation (a slow down of the growth of the structure 
functions) and the unitarization of the cross section. 

At very small $x$, the nucleon is a very dense system of gluons and can be 
described via the Color Glass Condensate formalism \cite{mv} which resums 
large logs of energy as well as the large gluon density effects. It reduces 
to the BFKL formalism \cite{bfkl} in the limit that the gluon density in a 
nucleon is small. The Color Glass Condensate is an all twist formalism and 
as such extends the domain of applicability of pQCD to high gluon density 
environments.

In this work, we consider different approaches to calculating the neutrino-nucleon total
cross section at ultra-high energies. First, we show the results from standard pQCD 
(DGLAP) \cite{dglap}  
approach as well as the results from a unified DGLAP/BFKL approach, available in the 
literature \cite{kk}. We then consider the neutrino-nucleon cross section using 
the Color Glass Condensate formalism and gluon saturation based approaches. This involves 
modeling the quark-anti quark dipole cross section which is the basic ingredient in the 
structure functions. We compare the resulting neutrino-nucleon cross sections from 
different approaches and comment on the possibility of using future neutrino observatories 
to constrain the ultra-high energy neutrino-nucleon cross sections.

\section{Neutrino-Nucleon Total Cross Section}
\subsection{Leading Twist pQCD}

In perturbative QCD (pQCD), the cross section for the neutrino nucleon cross section
can be written as
\be
\sigma_{total}^{\nu N}(s)=\int_0^1 dx \int_0^{xs} dQ^2
{d^2 \sigma^{\nu N} \over dxdQ^2}\, ,
\label{eq:stcs}
\ee
where the differential cross section is given in terms of the quark and anti-quark
distribution functions 
\be
{d^2 \sigma^{\nu N} \over dxdQ^2}={G_F^2 \over \pi}
\bigg({M^2_{W,Z} \over Q^2 + M^2_{W,Z}}\bigg)^2 
\bigg[q(x,Q^2) + (1- Q^2/xs)^2 {\bar q}(x,Q^2)\bigg].
\label{eq:difcs}
\ee
Here $G_F$ is the Fermi constant and $M_{W,Z}$ refer to the $W$ or $Z$ boson masses
while $s$ is the neutrino-nucleon center of mass energy. The total cross section
is finite (unlike the photon exchange process) and is dominated by scales 
$Q \sim M_{W,Z}$. In what follows, we will restrict ourselves to charged current
exchanges, but the extension of work to the case of neutral current is 
trivial and we expect our results for the charged current exchange to hold equally
well for the neutral current exchange. 

In the standard Leading Twist
(LT) pQCD approach, one parameterizes the $x$ dependence of quark and anti-quark 
distribution functions $q(x,Q^2), \bar{q}(x,Q^2)$ at some initial scale $Q_0$, 
typically taken 
to be of order of a GeV or so. The distribution functions are then given by DGLAP 
evolution equations at any other $x$ and $Q > Q_0$. The parameterizations are fit to 
the available data on DIS, for example, at HERA. There are
various parameterizations of parton distribution functions satisfying the DGLAP
evolution equations, for example CTEQ, MRST and GRV which differ in the choice of
initial conditions and the degree of sophistication. 

\begin{figure}[htp]
\centering
\setlength{\epsfxsize=10cm}
\centerline{\epsffile{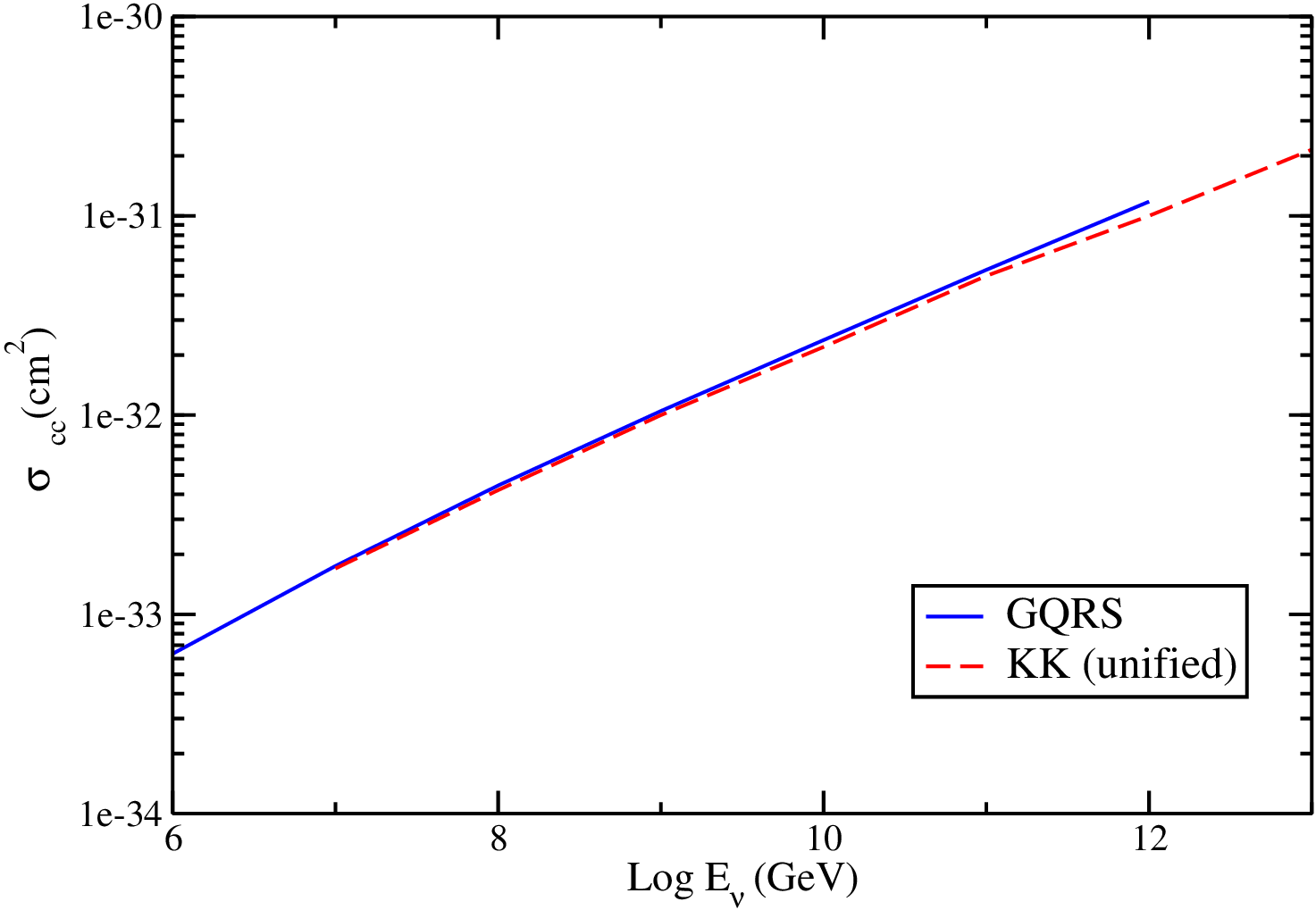}}
\caption{The neutrino-nucleon cross section in Leading Twist pQCD via charged current 
exchange \cite{reno,kk}.}
\label{fig:kk_gqrs}
\end{figure}

If the neutrino-nucleon center of mass energy is much higher than the exchanged
momentum scale such that  $\alpha_s \, \ln s/M_W^2 \sim 1$, it is more appropriate
to use the BFKL formalism which resums these large logs rather than the DGLAP
formalism. It is also possible to combine the two approaches in a phenomenological
way such that both DGLAP and BFKL resummations are included \cite{kk}. In 
Fig. (\ref{fig:kk_gqrs}) we show 
the results of a DGLAP based calculation of the neutrino-nucleon total cross section, 
via charged current exchange due to Gandhi et al. \cite{reno}, denoted GQRS,  
as well as a calculation due to Kutak et al., denoted KK (unified), which uses a 
unified DGLAP and BFKL approach (shown here without gluon saturation effects). 
The cross section grows with the center of mass 
energy which can be parameterized in terms of the incident neutrino energy (in 
the range shown in Fig. \ref{fig:kk_gqrs}) as $\sigma \sim (E_{\nu}/ 1 GeV)^{0.402}$. 
It can be shown that this increase in the cross section is due to the growth of the 
parton distribution functions with decreasing Bjorken $x$ \cite{dkrs}. While at the lowest energy 
the two results are identical, which shows small $x$ effects resummed by BFKL are 
negligible, at higher neutrino energy the two results can differ by a factor of two 
or larger. This signifies the fact that it is essential to include the contribution 
of small $x$ partons properly at ultra-high energies.

\vspace{0.3in}
It is important to realize that the HERA data on DIS covers a limited kinematic
region and that ultra-high energy neutrino-nucleon cross sections are dominated
by gluons at very small $x$ and high $Q^2$ where there is no data. In the standard
approach, one extrapolates the solution of the DGLAP evolution equations for parton 
distribution functions to smaller $x$, as needed. However, this requires making
assumptions (or rather educated guesses) about the behavior of the distribution
functions at small $x$. As we will show below, making rather plausible assumptions
about the behavior of the parton distribution functions at small $x$, leads to 
large variations of the cross section at ultra-high energies.

\subsection{Gluon Saturation}

At very small Bjorken $x$, the gluon distribution function is expected to saturate, 
which would lead to a slow down of the growth of the neutrino-nucleon total cross
section with energy. This is accomplished in the Color Glass Condensate (CGC) formalism
which is an effective theory of QCD at high energy. 
The differential neutrino-nucleon cross 
section can be written in terms of the structure functions $F_1$ and $F_2$ ($F_3$ does not 
contribute at small x),
\be
\frac{d^2\sigma}{dx dQ^2} = \frac{1}{2\pi} \frac{G_F^2}{(1+ Q^2/M_W^2)^2}
[(1-y) F_2 (x,Q^2) + y^2 x F_1 (x,Q^2)]
\label{eq:diff_cs}
\ee
with 
\be
F_2 &=& \frac{N_c Q^2}{4\pi^3}\int_0^1 dz \int dr_t^2  \sigma_d (x,r_t)
\{
4z^2(1-z)^2Q^2 K_0^2(ar_t) + a^2[z^2 + (1-z)^2]K_1^2(ar_t)
\}\nonumber \\
F_1 &=&\frac{1}{2x} \frac{N_c Q^2}{4 \pi^3} \int_0^1 dz \int dr_t^2  \sigma_d (x,r_t)
{a^2[z^2 + (1-z)^2]K_1^2(ar_t)}
\label{eq:f2_f1} 
\ee
where  $a^2 = z(1-z)Q^2$ and $K_0$ and $K_1$ are modified Bessel functions, 
$r_t$ is the size of the dipole and $z$ is the fraction of the photon energy carried
by the quark. The total  
cross section is the integral of (\ref{eq:diff_cs}) over $x$, from $x_{min} = Q^2/s$ 
to 1 and over $Q$, where we choose $Q_{min}$ to be $10  GeV$. The total cross section
does not receive any appreciable contribution from scales below $Q_{min}$. The
essential ingredient in saturation based approaches is the dipole cross section
which is the imaginary part of the forward scattering amplitude (hence the name dipole 
cross section) of a quark anti-quark dipole on the nucleon. The dipole cross section
$\sigma_d (x,r_t)$ satisfies the JIMWLK evolution equation \cite{jimwlk} which is the all twist 
generalization of the BFKL evolution equation. In practice since the JIMWLK evolution
equation is a highly non-linear equation, it is easier to parameterize the
dipole cross section, in analogy to parameterizations of the standard parton distribution
functions. The parameterizations of the dipole cross section are then used to 
calculate the structure functions in (\ref{eq:f2_f1}) and checked against available data 
in DIS \cite{gbw,bgbk,iim,lub}. The Color Glass Condensate formalism has also been successfully 
applied to particle production data in dA collisions at RHIC \cite{kkt,aaj} (for a 
review see \cite{iv}). The 
dipole cross section depends sensitively on the value of the saturation scale $Q_s$
and its energy dependence. While the overall magnitude of the saturation scale can not
be determined from CGC itself, its energy ($x$) dependence is computed from CGC 
itself \cite{dino} and is in good agreement with the value extracted from HERA 
phenomenology which has been parameterized \cite{gbw} as  
\be
Q_s^2(x) = (1 GeV^2) (3 \times 10^{-4}/x)^{.28} \;.
\label{eq:sat_scale}
\ee

The value of the saturation scale $Q_s$ compared to $M_W$ determines whether one 
is in the saturation region ($Q_s \ge M_W$), in the so called geometric scaling 
\cite{geo_sca}
region ($Q_s \le M_W \le Q_s^2/\Lambda_{QCD}$) or in the DGLAP region 
($ Q_s^2/\Lambda_{QCD} \le M_W$). It is ideal to have a unified formalism which 
can address all three regions; however, such a formalism does not exist
currently. One can either use the DGLAP evolution equation and modify it to 
include gluon saturation effects as in \cite{kk} or use the CGC formalism
and add the contributions of the DGLAP region by using the standard pQCD
expressions. We choose the later approach since we are mainly interested in 
the ultra-high energy neutrino cross sections where the main contribution
to the cross section comes from the very small $x$ region. To do this, we introduce 
a cutoff $x_0$ below which we use the CGC expressions (\ref{eq:f2_f1}) while for
$x > x_0$ we use (\ref{eq:difcs}) where the quark and anti-quark distributions are
taken from CTEQ parameterization. 

One of the earlier parameterizations of the dipole cross section is due to 
Bartels et al. \cite{bgbk} which has been used to fit the HERA data. It is given by
\be
\sigma_d (x, r_t) = \sigma_0
[1 - exp(\pi^2 r_t^2 \alpha_s(\mu^2) xg(x, \mu^2)/(3 \sigma_0))] 
\label{eq:cs_bgbk}
\ee
with $\mu^2 = .26/r_t^2 + 0.52$ and the gluon distribution function $xg$ satisfies
the DGLAP evolution equation. The overal constant $\sigma_0$ is the nucleon size and taken to be 
$\sigma_0=23 mb$. This parameterization includes higher twist effects but does not have 
the BFKL anomalous dimension. Another parameterization is due to Kharzeev et al. \cite{kkt}
and has been used to fit the RHIC data on deuteron-nucleus collisions \cite{kkt,aaj}.
The dipole cross section in this parameterization is given by
 \be 
\sigma_d (r_t, y) = \sigma_0 \, \left (\exp\left[ - \frac{1}{4} [r_t^2
  Q_s^2(y)]^{\gamma(y,r_t)}\right] -1 \right )
  \label{eq:cs_kkt}
\ee
where the saturation scale is given by $Q_s(y) = Q_0 \exp [\lambda (y-y_0)/2] $ 
with $y = \ln 1/x$ and $y_0=0.6, \, \lambda = 0.3$. The anomalous dimension $\gamma$ is 
\be
\gamma(y,r_t) = \frac{1}{2}\left(1+
\frac{\xi(y,r_t)}{\xi(y,r_t)+\sqrt{2\xi(y,r_t)}+28\zeta(3)}
   \right)
   \label{eq:ano_kkt}
\ee
where
\be
\xi (y,r_t) = \frac{\log 1/r_t^2 Q_0^2}{(\lambda/2)(y-y_0)}~.
\ee
This parameterization has the advantage that, unlike the one in (\ref{eq:cs_bgbk}), 
it has the BFKL anomalous dimension built in which seems to be essential in describing
the forward rapidity deuteron-gold data at RHIC. Using these two
parameterizations of the dipole cross section, we calculate the neutrino-nucleon 
total cross section. We assume that quark (anti-quark) distributions are known well  
for $x \geq x_0$ and use (\ref{eq:difcs}) to calculate the cross section for $x \geq x_0$.
For $x \leq x_0$, we use the saturation approach and calculate the cross section
using (\ref{eq:diff_cs}) with the structure functions given by (\ref{eq:f2_f1}), using the 
two different parameterizations of the dipole cross 
section given in (\ref{eq:cs_bgbk}, \ref{eq:cs_kkt}), denoted BGBK and KKT dipoles respectively.
To check the sensitivity of our results to the choice of $x_0$, we try two different values of
$x_0$, first $x_0 = 10^{-4}$ and then $x_0 = 10^{-6}$.
In case of BGBK dipoles, since gluon distribution function $xg (x, \mu^2)$ is not known 
well below $x \leq 10^{-5}$,
we consider three wildly different scenarios; (i) a continually growing 
distributions for $x \leq 10^{-5}$, (ii) a flat distribution for $x \leq 10^{-5}$,
and (iii) a distribution which falls by one order of magnitude for every decade of 
decreasing x for $x\leq 10^{-5}$. A measurement of the neutrino-nucleon cross sections 
at ultra high energies would thus go a long way toward understanding the very small x 
parton distributions.

\section{Results and Discussion}  

\begin{figure}[htp]
\centering
\setlength{\epsfxsize=10cm}
\centerline{\epsffile{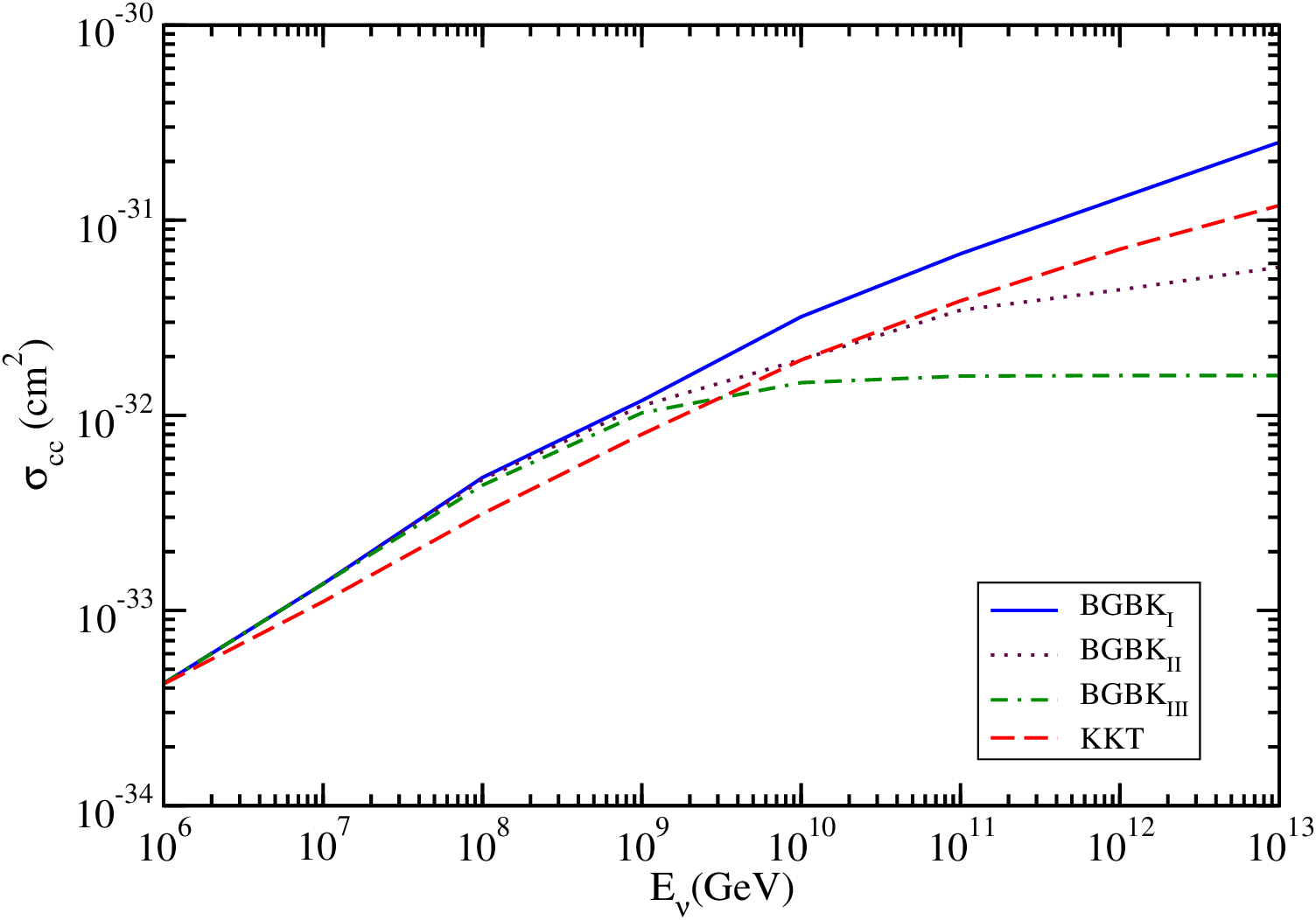}}
\caption{The neutrino-nucleon cross section with $x_0=10^{-4}$, details are given in the text.}
\label{fig:cs_4}
\end{figure}

In Fig. (\ref{fig:cs_4}) we show our results for the neutrino-nucleon total cross section
(via charged current exchange) for different neutrino energies for the case where 
$x_0=10^{-4}$ and BGBK denotes the Bartels et al. model of the dipole cross section
given by (\ref{eq:cs_bgbk}) and KKT denotes the Kharzeev et al. parameterization 
given in (\ref{eq:cs_kkt}). The subscript $I$ refers to the case where the gluon
distribution function $xg (x, \mu^2)$ in (\ref{eq:cs_bgbk}) , taken from CTEQ6, 
keeps growing with $x$ below $x= 10^{-5}$
while $II$ refers to the case where the gluon distribution function below $x=10^{-5}$
is flat and finally, case $III$ corresponds to the case where the gluon distribution
function below $x=10^{-5}$ falls like a power.

\vspace{0.3in}

For neutrino energies less than $10^8$ GeV, the cross section does not receive significant
contributions from the region where $x < 10^{-5}$. This shows in Fig. (\ref{fig:cs_4}) as 
the three cases $I, II, III$ (Bartels et al. dipole, denoted BGBK, with different gluon 
behavior at small $x$)
being almost identical for $E_{\nu} < 10^8$ GeV while the cross section calculated using 
the KKT parameterization of the dipole profile starts out below the other dipole models
until about neutrino energy of $10^8 - 10^9$ GeV after which it passes the BGBK $II,III$ 
dipoles, due to the constancy or drop off of the BGBK gluon distribution function below
$x=10^{-5}$.
\vspace{0.3in}

\begin{figure}[hbp]
\centering
\setlength{\epsfxsize=10cm}
\centerline{\epsffile{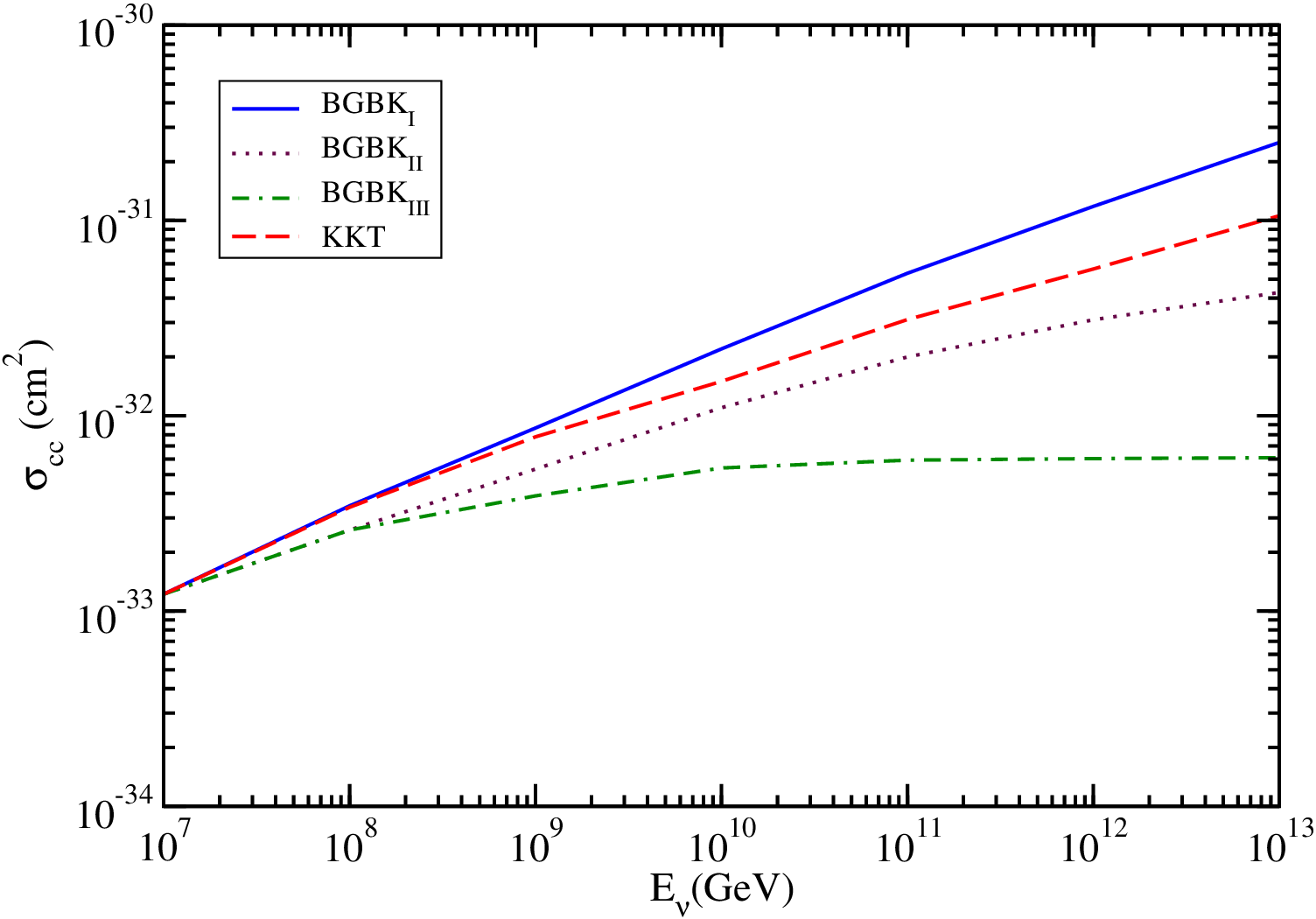}}
\caption{The same as in Fig. (\ref{fig:cs_4}) but with $x_0=10^{-6}$.}
\label{fig:cs_6}
\end{figure}
To see the sensitivity of our results to the choice of cutoff $x_0$, we show the 
neutrino-nucleon cross section in Fig. (\ref{fig:cs_6}) with the cutoff $x_0$
now taken to be $10^{-6}$. Again, for $x > x_0$, we use the quark and anti-quark distribution
functions in (\ref{eq:difcs}) to calculate the cross section while for the region $x < x_0$
we use the saturation approach. While the BGBK $I$ does not change as it must not, the case where
we have the gluon distribution function falling off (BGBK $III$) is severely affected, 
by as much as a factor of $4$ at the highest energy shown. On the other hand, the cross section
using the KKT parametrization is rather robust, a change in $x_0$ from $10^{-4}$ to $10^{-6}$
changes the cross section by about $20 \%$ at $E_{\nu} = 10^{10}$ and $10 \%$ at 
$E_{\nu} = 10^{13}$.
Depending on $x_0$, the cross section given by Gandhi et al. \cite{reno} is about 
$1.65 -2.0$ times bigger than the KKT cross section at $E_{\nu} = 10^{12}$.
It is clear that the assumptions made on the behavior of the gluon distribution function at very
small $x$ will determine the outcome of the calculated cross sections at high energy.
\vspace{0.3in}

\begin{figure}[hbp]
\centering
\setlength{\epsfxsize=10cm}
\centerline{\epsffile{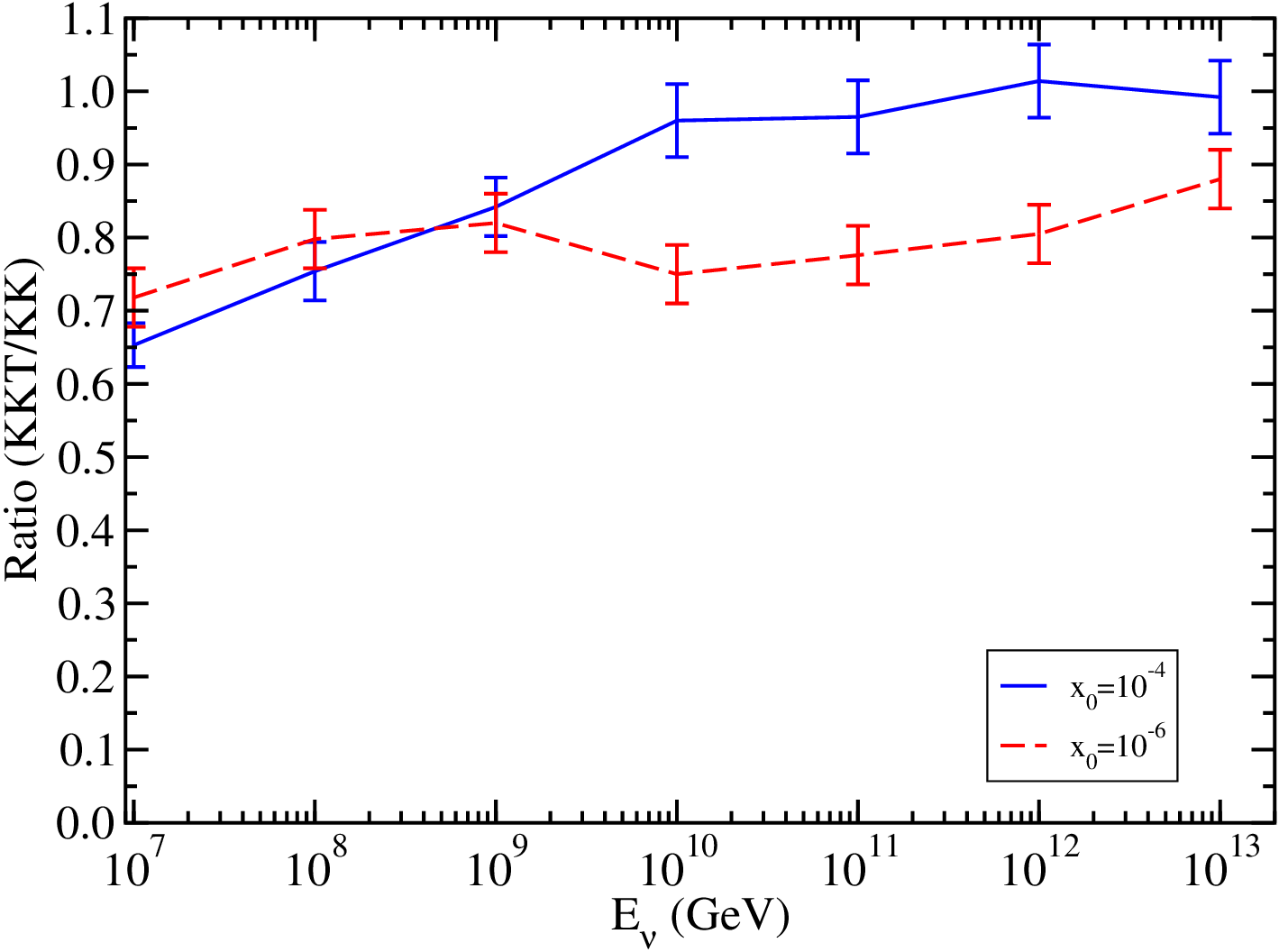}}
\caption{Ratio of KKT and KK (screened) cross sections .}
\label{fig:r_kkt-kk_s}
\end{figure}
To compare our results to other saturation motivated studies, we show the ratio of our
results for the neutrino-nucleon cross section using the KKT paraneterization, denoted KKT 
and the results of (screened) Kutak and Kwiecinski \cite{kk}, denoted KK for the two choices of the
parameter $x_0$ in Fig. (\ref{fig:r_kkt-kk_s}). Since the numerical integrations involved
are quite time consuming, we have have taken rather large increments in the integration 
rountines which
leads to about $10\%$ error on the KKT cross sections. This is the origin of the error bars
shown in the figure. For neutrino energies more than $10^8 - 10^9$, the two approaches are in 
exccelent agreement for $x_0=10^{-4}$ and within $20 \%$ for $x=10^{-6}$. The agreement is rather
remarkable since the KK approach involves solving a phenomenologically unified DGLAP/BFK 
equation with a non-linear term motivated by the saturation physics while the results denoted 
KKT are based on a parameterization of the dipole profile which is motivated by the RHIC data 
on deuteron-gold collisions \cite{iv}. This is most likely due to the similar growth of the
saturaion scale in both cases since this growth is measures at HERA. It is also calculated
very reliabely in the Color Glass Condensate formalism \cite{dino} and is in excellent 
agreement with
the measured value at HERA. The fact that the two rather different approaches give 
quite similar results for neutrino-nucleon cross section at high energies is very reassuring
and gives us confidence that if and when the ultra high energy neutrino-nucleon cross sections
are measured, one can have quite stringent constraints on saturation based calculations of the  
neutrino-nucleon cross section.

\section{Acknowledgments}
The authors thank Mary Alberg and  William Detmold for some computing help and
W-Y. P. Hwang for earlier collaboration on this work. E. H. is supported in part 
by the U.S. \ Department of Energy under Grant No.\ DE-FG03-97ER4014. 
J. J-M. is supported in part by the U.S. \ Department of Energy under 
Grant No.\ DE-FG02-00ER41132.

 .

\end{document}